\documentclass[aps,pre,twocolumn,groupedaddress,showpacs,amsmath]{revtex4}
    \usepackage{bm}
    \usepackage{graphicx}
\newcommand\beq{\begin{equation}}
\newcommand\eeq{\end{equation}}
\newcommand\beqa{\begin{eqnarray}}
\newcommand\eeqa{\end{eqnarray}}

\begin{document}
\title{Structure of ternary additive hard-sphere fluid
mixtures }
\author{Alexander Malijevsk\'{y}}
\email[]{Alexandr.Malijevsky@vscht.cz}
\author{Anatol Malijevsk\'{y}}
\email[]{Anatol.Malijevsky@vscht.cz} \affiliation{Institute of
Chemical Technology, 16628 Prague 6, Czech Republic}
\author{Santos B. Yuste}
\email[]{santos@unex.es}
\author{Andr\'es Santos}
\email[]{andres@unex.es}
\affiliation{Departmento de
F\'{\i}sica, Universidad de Extremadura,
Badajoz, E-06071, Spain}
\author{Mariano L\'{o}pez de Haro}
\email[]{malopez@servidor.unam.mx}
\affiliation{Centro de Investigaci\'on en
Energ\'{\i}a, UNAM, Temixco, Morelos 62580, M{e}xico}
\date{\today}

\begin{abstract}
Monte Carlo simulations on the structural  properties of
ternary fluid mixtures of additive hard
spheres are reported. The results are compared with those obtained from a
recent analytical approximation [S. B. Yuste,
A. Santos, and M. L\'{o}pez de Haro, J. Chem. Phys. {\bf 108}, 3683 (1998)]
to the radial distribution functions of hard-sphere mixtures and with the
results derived from the solution of the
Ornstein--Zernike integral equation with both the
Martynov--Sarkisov and the Percus--Yevick closures.
 Very good agreement
between the results of the first two approaches and simulation is
observed, with a noticeable improvement over the Percus--Yevick
predictions especially near contact.
\end{abstract}
\maketitle

\section{Introduction\label{sect1}}

It is widely recognized that hard-sphere fluids have played a key
role in the development and consolidation of liquid state theory.
For these model systems, the link between structural properties
and thermodynamics is immediate and simple, leading to rather
straightforward expressions for the internal energy (which reduces
to that of the ideal gas), and for the pressure equation, which
only involves the contact values of the radial distribution
functions (rdf)\cite{McQuarrie,BNH80,RS82}. Nevertheless and
despite the vast amount of literature devoted to their study, up
to this day even the derivation of an explicit (exact) equation of
state for these systems remains as an open problem. Under these
circumstances, computer simulations have proved to be a useful way
to derive structural and thermodynamic information as well as to
allow the assessment of the many approximate theories proposed for
them. These theories range from useful empirical expressions for
the contact values of the rdf or the equation of state to the
solution of Ornstein--Zernike (OZ) integral equations with a given
closure. And of course both theory and simulation increase their
complexity if one considers mixtures rather than single component
fluids, so that it is not surprising that the available results
are much scarcer for hard-sphere mixtures than for pure
hard-sphere fluids. In fact, only binary mixtures have received
some attention while results for ternary hard-sphere mixtures and
those composed of more than three components are particularly
limited. As far as we are aware, there is only one computer
simulation study on the structure and thermodynamics of the
hard-sphere additive ternary mixture
 \cite{bou2}, where diameter ratios
1:2:$\frac{10}{3}$ were considered at three densities and two
compositions. One should also mention that Schaink \cite{Schaink}
has performed a simulation study of a \textit{non-additive}
hard-sphere ternary mixture where the diameter ratios 1:1:1 were
taken; {the same mixture was studied by Gazzillo \cite{G95} from
an integral equation approach}. On the theoretical side, it is
imperative to mention the pioneering work of Lebowitz  
\cite{lebowitz,PRK98},  who
solved the Percus--Yevick equation for a multicomponent mixture of
additive hard spheres. Also important are the papers by
Boubl\'{\i}k \cite{B70}, Grundke and Henderson \cite{GH72}, and
Lee and Levesque \cite{LL73},
 in
which they introduced the contact values, now referred to as the
Boubl\'{\i}k--Grundke--Henderson--Lee--Levesque (BGHLL) contact
values, leading to the
Boubl\'{\i}k--Mansoori--Carnahan--Starling--Leland (BMCSL)
equation of state \cite {B70,MCSL}
Apart from these, in the case of multicomponent mixtures to our
knowledge there is only
some work by Gazzillo
\cite{gazzillo} on the thermodynamic criteria of local stability,
a paper by Boubl\'{\i}k \cite{bou1} on rdf, the scaled field
particle theory of isotropic hard particle fluids of Rosenfeld
\cite{R88} and the studies carried out by some of us
\cite{BSL98,Getafe}. In these latter
studies an interesting behavior of the rdf
$g_{ij}(r)$ was predicted, but it could not be assessed in view of the
then absence of available computer simulation data to compare
with.

On another vein, it is clear that ternary mixtures are typical in nature and
technology. For instance, air is essentially
a mixture of nitrogen, oxygen, and argon (the concentration of other
components is much lower), and sea water is a mixture
of H$_{2}$O, Na$^{+}$ and Cl$^{-}$. There is also a number of industrially
important chemical reactions among three
components, {\it e.g.} the synthesis of ammonia
\[
3\text{H}_{2}+\text{N}_{2}=2\text{NH}_{3},
\]
or in ecology, {\it e.g.}
\[
\text{SO}_{2}+\frac{1}{2}\text{O}_{2}=\text{SO}_{3}.
\]
{Ternary mixtures of molecules whose interaction includes an
attractive part have been studied from perturbation theory and van der
Waals one-fluid theory \cite{FS96,BD99,WS99}.}

In view of the above, the aim of this paper is to provide
simulation results for hard-sphere additive ternary mixtures that
will serve as a starting point to assess the accuracy and validity
of some theoretical approaches. Specifically, we will examine five
ternary systems at the same packing fraction and with fixed
diameter ratios, so that they are only different because of their
composition. Two of these cases correspond to mixtures in which
the biggest spheres occupy over 50\% of the available volume,
followed in volume occupation by the intermediate sized spheres
and finally by the smallest spheres. A third system is considered
in which all species share equitatively the available volume,
while in the last two systems it is the {intermediate} spheres
which occupy the {smallest} volume and either the biggest or the
{smallest} sized ones follow in volume occupation. The theoretical
approaches that we will consider will be the solution of the
Ornstein--Zernike equation with both the Percus--Yevick
\cite{lebowitz} and the Martynov--Sarkisov  \cite{MS} closures and
the (approximate) expressions for the rdf of a hard-sphere mixture
derived in Ref.\ \cite{BSL98}.

The paper is organized as follows. In order to make the paper
self-contained, in Sect.\ \ref{sect2} we recall the main
results of the theoretical approaches to derive the structural properties of
hard-sphere mixtures. Section \ref{sect3} provides some
details of the simulation and the comparison between simulation data and the
different theoretical approximations. We
close the paper in Sect.\ \ref{sect4} with a discussion and some concluding
remarks.

\section{Theoretical Approximations to the Structural Properties
of {Multicomponent} Mixtures of Additive Hard
Spheres\label{sect2}}

An $n$-component mixture made of $\rho _{i}$ hard spheres (of
diameter $ \sigma _{i}$) per volume unit may be characterized by
 $2n-1$ parameters (for
instance, the $n-1$ mole fractions $x_{i}=\rho_i/\rho$, the $n-1$
size ratios $\sigma _{i}/\sigma _{1}$ and the packing fraction
$\eta \equiv \sum_{i=1}^{n}\eta _{i}$, $\eta _{i}=\frac{\pi
}{6}\rho_{i}\sigma _{i}^{3}$ denoting the partial packing fraction
corresponding to species $i$) and involves $n(n+1)/2$ rdf
$g_{ij}(r)$. Within the usual integral equation approach, the OZ
equation is a set of $n(n+1)/2$ coupled equations
\begin{equation}
\gamma _{ij}(r)=\rho \sum_{k=1}^nx_k \int d{\bf r^{
\prime }}\,h_{ik}(|{\bf r^{\prime }}|)
c_{kj}(|{\bf r-r^{\prime }}|),
 \label{OZ}
 \end{equation}
 where $h_{ij}(r)\equiv g_{ij}(r)-1$ and
$c_{ij}$ denote the total and the direct
correlation functions, respectively, and $\gamma _{ij}=h_{ij}-c_{ij}$ is the
series function. A general closure for the
OZ equation may be written in the form
\begin{equation}
c_{ij}(r)=\exp \left
[ -\beta u_{ij}(r)+\gamma _{ij}(r)+B_{ij}(r)
\right] -1-\gamma _{ij}(r),
\label{cl}
\end{equation}
where $u_{ij}(r)$ is the interaction potential and $B_{ij}$ is
the bridge function. In this work we consider
two approximations to the bridge function: the classical Percus--Yevick (PY)
theory
\begin{equation}
B_{ij}(r)=\ln [1+\gamma _{ij}(r)]-\gamma_{ij}(r),
\label{PY}
\end{equation}
and the Martynov--Sarkisov (MS) \cite{MS} theory
\begin{equation}
B_{ij}(r)=\sqrt{1+2\gamma _{ij}(r)}-1-\gamma _{ij}(r).
\label{MS}
\end{equation}

We solved the OZ equation with these closures {in the case of ternary
mixtures ($n=3$)} using an algorithm that is a
direct extension of the method proposed for
one-component simple fluids \cite{lab}. In our numerical implementation in
this paper, we used $N=2048$ grid points
with a step size $\Delta r=0.01$. {In the case of the PY closure, the
rdf can also be obtained by numerical inversion of  analytical
expressions in Laplace space \cite{BSL98}. Both methods give
undistinguishable results, what gives confidence on the accuracy of the
numerical solution of the MS closure.}

An alternative method to obtain an approximate expression for
$g_{ij}(r)$ for a multicomponent mixture was introduced in Ref.
\cite{BSL98}. We will refer to this method as the RFA approach
since it stemmed out of a generalization of a rational function
approximation to structural quantities in a simple hard-sphere
fluid \cite{BS91}. Working in the Laplace space and defining
$G_{ij}(s)=\int_{0}^{\infty }dr\,e^{-sr}rg_{ij}(r)$,  the
foregoing approach implies that $ G_{ij}$ is assumed to adopt the
following functional form \cite{BSL98,YSH00}:
\begin{equation}
G_{ij}(s)=\frac{e^{-s\sigma _{ij}}}{2\pi s^{2}}\sum_{k=1}^{
n}L_{ik}(s){ [(1+\alpha s){\sf I}-{\sf A}(s)]
^{-1}}_{kj},
\label{G(s)}
\end{equation}
 where \beq
L_{ij}(s)=L_{ij}^{(0)}+L_{ij}^{(1)}s+L_{ij}^{(2)}s^{2},
\label{new1} \eeq \beq A_{ij}(s)=\rho _{i}\sum_{p=0}^{2}\varphi
_{p}(s\sigma _{i})\sigma _{i}^{p+1}L_{ij}^{(2-p)}, \label{new2}
\eeq
 with
\beq
\varphi _{p}(x)\equiv
x^{-(p+1)}\left[\sum_{m=0}^{p}\frac{(-x)^{m}}{m!}-e^{-x}\right].
\label{new3}
\eeq
 There are two basic
requirements that $G_{ij}(s)$ must satisfy. First, since
$g_{ij}(r)=0$ for $r<\sigma _{ij}$, with $\sigma _{ij}=\left(
\sigma _{i}\ +\sigma _{j}\right) /2$, and the contact values
$g_{ij}(\sigma _{ij}^{+})= \mbox{finite}$, this implies that (i)
$\lim_{s\rightarrow \infty }s\,e^{s\sigma
_{ij}}G_{ij}(s)=\mbox{finite}$. Second, the isothermal
compressibility $\kappa _{T}=\mbox{finite}$, so that (ii) $
\lim_{s\rightarrow 0}[G_{ij}(s)-s^{-2}]=\mbox{finite}$. Condition
(i) is verified by construction. On the other hand, condition (ii)
yields two {\em linear\/} sets of $n^2$ equations each, whose
solution is straightforward:
\beq
 L_{ij}^{(0)}=\lambda +\lambda
^{\prime }\sigma _{j}+2\lambda ^{\prime }\alpha -\lambda
\sum_{k=1}^{n}\rho _{k}\sigma _{k}L_{kj}^{(2)}, \label{new4} \eeq
\beq L_{ij}^{(1)}=\lambda \sigma _{ij}+\frac{\lambda ^{\prime
}}{2}\sigma _{i}\sigma _{j}+(\lambda +\lambda ^{\prime }\sigma
_{i})\alpha -\frac{ \lambda }{2}\sigma _{i}\sum_{k=1}^{n}\rho
_{k}\sigma _{k}L_{kj}^{(2)}, \label{new5} \eeq where $\lambda
\equiv 2\pi /(1-\eta )$ and $\lambda ^{\prime }\equiv (\lambda
/2)^{2}\rho \langle \sigma^2\rangle$ with $\langle \sigma
^{p}\rangle \equiv \sum_{i=1}^{n}x_{i}\sigma _{i}^{p}$. The
parameters $ L_{ij}^{(2)}$ and $\alpha $ are arbitrary, so that
conditions (i) y (ii) are satisfied regardless of their choice. In
particular, if one chooses $ L_{ij}^{(2)}=\alpha =0$, the
approximation given by Eq. (\ref{G(s)}) coincides with the PY
solution. If, on the other hand, we fix given values for
$g_{ij}(\sigma _{ij}^{+})$, we get the relationship
$L_{ij}^{(2)}=2\pi \alpha \sigma _{ij}g_{ij}(\sigma _{ij}^{+})$;
thus, only $\alpha $ remains to be determined. Finally, if we fix
$\kappa _{T}$, we obtain an algebraic equation for $\alpha $ of
degree $2n$.

In previous work with the RFA approach \cite{BSL98,YSH00} the
BGHLL values of $g_{ij}(\sigma _{ij}^{+})$ and $\kappa _{T}$ given
by the BMCLS equation of state \cite{B70,MCSL} were considered. In
this work, however, we will use a different approximation which
was recently proposed by three of us \cite{SBL02}. Following this
proposal, we assume that
\begin{equation}
g_{ij}(\sigma _{ij}^{+})=F(\eta ,z_{ij}),
\label{contact}
\end{equation}
where
$z_{ij}\equiv (\sigma _{i}\sigma _{j}/\sigma _{ij})\langle \sigma
^{2}\rangle /\langle \sigma ^{3}\rangle $,  and
take the function $F(\eta ,z) $ to be {\it
universal\/} in the sense that it is a common function for all the pairs
$ij$.
Further $F$ is forced to comply with known
exact relations in the point particle, equal size and colloidal limits.
Under these circumstances, the simplest
functional form that $F$ may adopt is a quadratic function of $z$:
\begin{equation}
F(\eta ,z)=F_{0}(\eta )+F_{1}(\eta )z
+F_{2}(\eta )z^{2},
\label{10}
\end{equation}
where the coefficients are
explicitly given by
\begin{equation}
F_{0}(\eta)=\frac{1}{1-\eta },
\label{11a}
\end{equation}
\begin{equation}
F_{1}(\eta)=2(1-\eta )g(\sigma^{+})-\frac{2-\eta/2}{1-\eta } ,
\label{11b}
\end{equation}
\begin{equation}
F_{2}(\eta )=
\frac{1-\eta/2}{1-\eta }-(1-2\eta )g(\sigma ^{+}) .
\label{11c}
\end{equation}
Here, $g(\sigma ^{+})$ denotes the contact value of the radial distribution
function of a simple hard-sphere fluid. For
this latter, we take the one corresponding to the Carnahan-Starling equation
of state \cite{CS69}, namely
$g_{\text{CS}}(\sigma ^{+})=(1-\eta /2)/(1-\eta )^{3}$. With such choice,
Eqs.\ (\ref{contact}) and (\ref{10}) become
\begin{equation}
g_{ij}(\sigma_{ij}^+)=\frac{1}{1-\eta }+\frac{3}{2}\frac{\eta
(1-\eta /3)}{(1-\eta )^{2} }z_{ij}+\frac{\eta ^{2}(1-\eta
/2)}{(1-\eta)^{3}}z_{ij}^{2},
\label{16}
\end{equation}
and
the compressibility factor for the {mixture}, from
which $\kappa _{T}$ may be readily derived, reads
\begin{equation}
Z(\eta )=Z_{\text{BMCSL}}(\eta )-\frac{\eta ^
{3}}{(1-\eta )^{2}} \frac{\langle \sigma ^{2}
\rangle }{\langle \sigma ^{3}\rangle ^{2}}\left( \langle \sigma \rangle
\langle \sigma ^{3}\rangle -\langle \sigma ^{2}
\rangle ^{2}\right) ,
\label{ZHSTM}
\end{equation}
where the
compressibility factor associated with the BMCSL equation of
state \cite{B70,MCSL} is given in the present notation by
\begin{equation}
Z_{\text{BMCSL}}(\eta )=\frac{1}{1-\eta }+ \frac{3\eta \langle
\sigma \rangle \langle \sigma ^{2}\rangle }{(1-\eta )^{2}\langle
\sigma ^{3}\rangle }+\frac{ \eta ^{2}(3-\eta )\langle \sigma
^{2}\rangle ^{3}}{(1-\eta )^{3}\langle \sigma ^{ 3}\rangle ^{2}}.
\label{BMCSL}
\end{equation}
 {Equation (\ref{16}) represents in general a
significant improvement over the BGHLL contact values
\cite{SBL02}. On the other hand, the BMCSL equation of state
(\ref{BMCSL}) performs slightly better than  Eq.\ (\ref{ZHSTM}).
Although the RFA can be implemented by making any choice for
$g_{ij}(\sigma_{ij}^+)$ and $\kappa_T$, here we have taken, in
addition to  the contact values (\ref{16}), the isothermal
compressibility associated with Eq.\ (\ref{ZHSTM}) in order to
enforce thermodynamic consistency.}
\section{Simulation
Details and Results\label{sect3}} We used the standard NVT-Monte
Carlo method with periodic boundary conditions, employing a cell
index algorithm with six different cell sizes corresponding to a
number of interactions. The simulation cubic box contained
$N=2700$ particles in each case but one (case C), where $N=6777$
particles were used.

The initial system with no overlaps was generated by random
insertion of particles to an originally empty box. The sequence we
used is the following: the largest particles were inserted first
and the smallest ones at the end. Particles were mixed during this
procedure.  Starting with this initial
configuration, we generated the Monte Carlo chain as follows. The
acceptance ratio of trial moves was adjusted to 10--15\% for all
the components. Each run was divided into 21 blocks, each of which
included about 10$^{9}$ of the equilibrium configurations
generated and contained 300--500 analyses of the calculation of
the rdf $g_{ij}(r)$ in the whole range of 1200 intervals $r_{i}\pm
\Delta r/2$ (where the step size was $\Delta r=5\times
10^{-3}\sigma _{1}$) up to a distance $6\sigma _{1}$. The analysis
was performed after $1000$ trial moves per particle (more
precisely after $1000N$ trial moves of a randomly chosen
particle). The first block was then discarded and the next 20 were
used to sample the configuration space, calculate mean values for
the entire run and estimate the errors.

The systems we examined  had the same
packing fraction $\eta =0.49$ and fixed diameter ratios $\sigma
_{2}/\sigma _{1}=2$ and $\sigma _{3}/\sigma _{1}=3$ (for
convenience and without loss of generality we have chosen the
value of the diameter of the smallest spheres to be always $1$) so
that their only difference lies in the composition. They are
identified as
\begin{description}
\item[(A)]
 $x_{1}=0.7$, $x_{2}=0.2$, $x_{3}=0.1$,\\
${\eta_1}/{\eta}= 0.14$,
${\eta_2}/{\eta}= 0.32$,
${\eta_3}/{\eta}= 0.54$,
\item[(B)]  $x_{1}=0.6$, $x_{2}=0.2$, $x_{3}=0.2$,
\\
${\eta_1}/{\eta}\simeq 0.08$,
${\eta_2}/{\eta}\simeq 0.21$,
${\eta_3}/{\eta}\simeq 0.71$,
\item[(C)]  $x_{1}=\frac{216}{251}$, $x_{2}=\frac{27}{251}$,
$x_{3}=\frac{8}{251}$,
\\
${\eta_1}/{\eta}={\eta_2}/{\eta}={\eta_3}/{\eta}=\frac{1}{3}$,
\item[(D)]  $x_{1}=0.85$, $x_{2}=0.05$, $x_{3}=0.10$,
\\
${\eta_1}/{\eta}\simeq 0.22$,
${\eta_2}/{\eta}\simeq 0.10$,
${\eta_3}/{\eta}\simeq 0.68$,
\item[(E)]  $x_{1}=0.90$, $x_{2}=0.07$, $x_{3}=0.03$,
\\
${\eta_1}/{\eta}\simeq 0.396$,
${\eta_2}/{\eta}\simeq 0.247$,
${\eta_3}/{\eta}\simeq 0.357$.
\end{description}

\begin{figure}[tbp]
\includegraphics[width=.90 \columnwidth]
{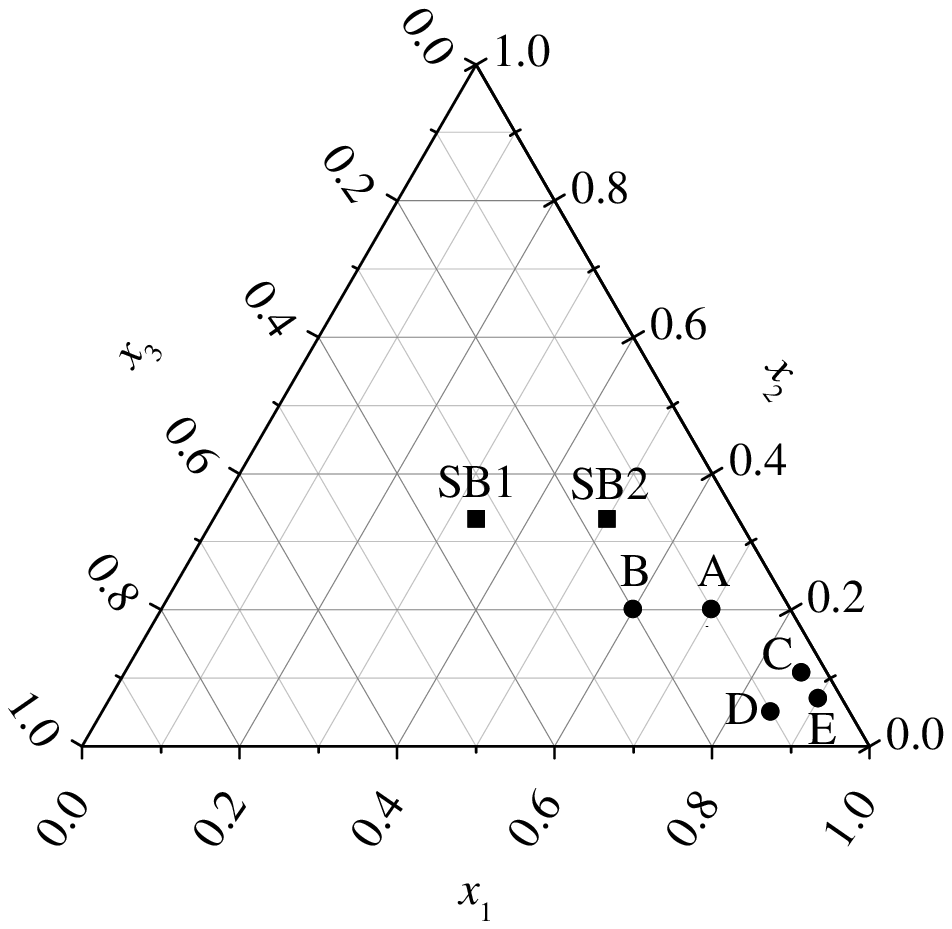}
\caption{{Ternary diagram showing the mole fractions of the five
cases A--E considered in this paper, as well as the two cases (SB1 and
SB2) considered by \v{S}indelka and Boubl\'{\i}k \protect\cite{bou2}.}
\label{fig1}}
\end{figure}

\begin{figure}[tbp]
\includegraphics[width=.90 \columnwidth]
{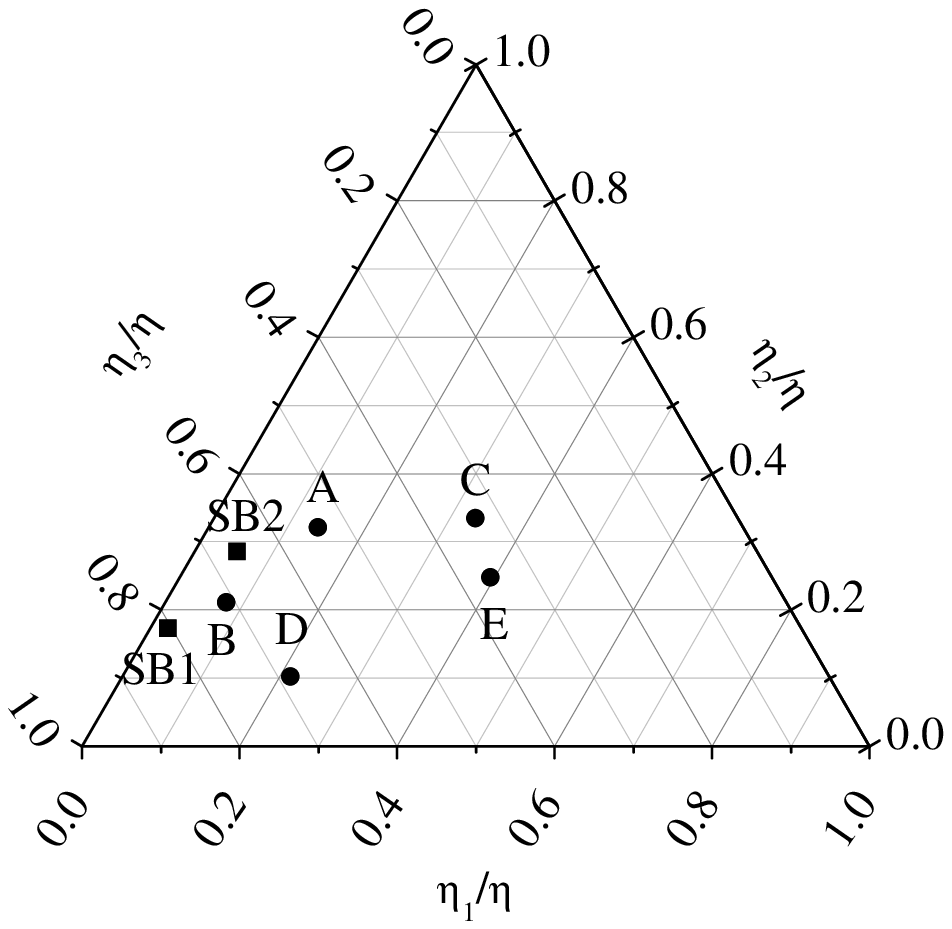}
\caption{{Ternary diagram showing the (relative) packing fractions of
the five cases A--E considered in this paper, as well as the two cases (SB1
and SB2) considered by \v{S}indelka and Boubl\'{\i}k \protect\cite{bou2}.}
\label{fig2}}
\end{figure}

These systems have been located in two different ternary diagrams,
one with respect to mole fractions and the other one corresponding
to partial packing fractions, shown in Figs.\ \ref{fig1} and
\ref{fig2}, respectively. In these diagrams we have also included
the two systems with  diameter ratios $\sigma _{2}/\sigma _{1}=2$
and $\sigma _{3}/\sigma _{1}=10/3$ that were studied by
\v{S}indelka and Boubl\'{\i}k \cite {bou2} and which we have
labeled SB1 ($x_1=x_2=x_3=\frac{1}{3}$; $\eta_1/\eta\simeq 0.022$,
$\eta_2/\eta\simeq 0.174$, $\eta_3/\eta\simeq 0.804$) and SB2
($x_1=\frac{1}{2}$, $x_2=\frac{1}{3}$, $x_3=\frac{1}{6}$;
$\eta_1/\eta\simeq 0.054$, $\eta_2/\eta\simeq 0.285$,
$\eta_3/\eta\simeq 0.661$). It should be pointed out that cases A
and B (as well as the systems SB1 and SB2) correspond to the
situation $\eta _{1}<\eta _{2}<\eta _{3}$, while in case D one has
$ \eta _{2}<\eta _{1}<\eta _{3}$, in case E $\eta _{2}<\eta
_{3}<\eta _{1}$ and in case C $\eta _{1}=\eta _{2}= \eta _{3}$.
This, in our view, allows us to examine the very different
situations that arise depending on which species occupies the
largest volume.

The results of our calculations, both theoretical and from the
simulations, are displayed in Figs.\ \ref{fig3}--\ref{fig8}. In
Fig.\ \ref{fig3} we show the contact values for all five systems
as functions of the parameter $z_{ij}$ defined below
Eq.~(\ref{contact}). In this instance we have considered the {PY,
BGHLL, MS, and scaled particle theory (SPT) values
\cite{LHP65,R88}, as well as those given by Eq.\ (\ref{16}). In
the case of the MS approximation we actually get a set of points
that have been joined by a line interrupted at $z_{33}=1.481$
(case D) since there is no convergence in cases C and E. The fact
that this line is a smooth one shows that the numerical values
obtained from the MS approximation seem to be consistent with the
``universality'' assumption (\ref{contact}).
 The comparison
with the simulation data indicates that for
$z>1$ the new proposal, Eq.\  (\ref{16}), improves over
the BGHLL prescription (while for $z<1$
it is only slightly worse) and both are
clearly superior to the SPT recipe. {The PY values are very poor,
while the MS approximation tends to underestimate the contact values for
$z>1$.} This provides some support to the use
of  Eq.\  (\ref{16}) and
Eq.\ (\ref{ZHSTM}) (this latter to compute $\kappa _{T}$) within the
approximate scheme to derive the rdf $ g_{ij}(r)$
for ternary mixtures that was introduced in Ref.\ \cite{BSL98}  and briefly
sketched in
the previous section.
\begin{figure}[tbp]
\includegraphics[width=.90 \columnwidth]
{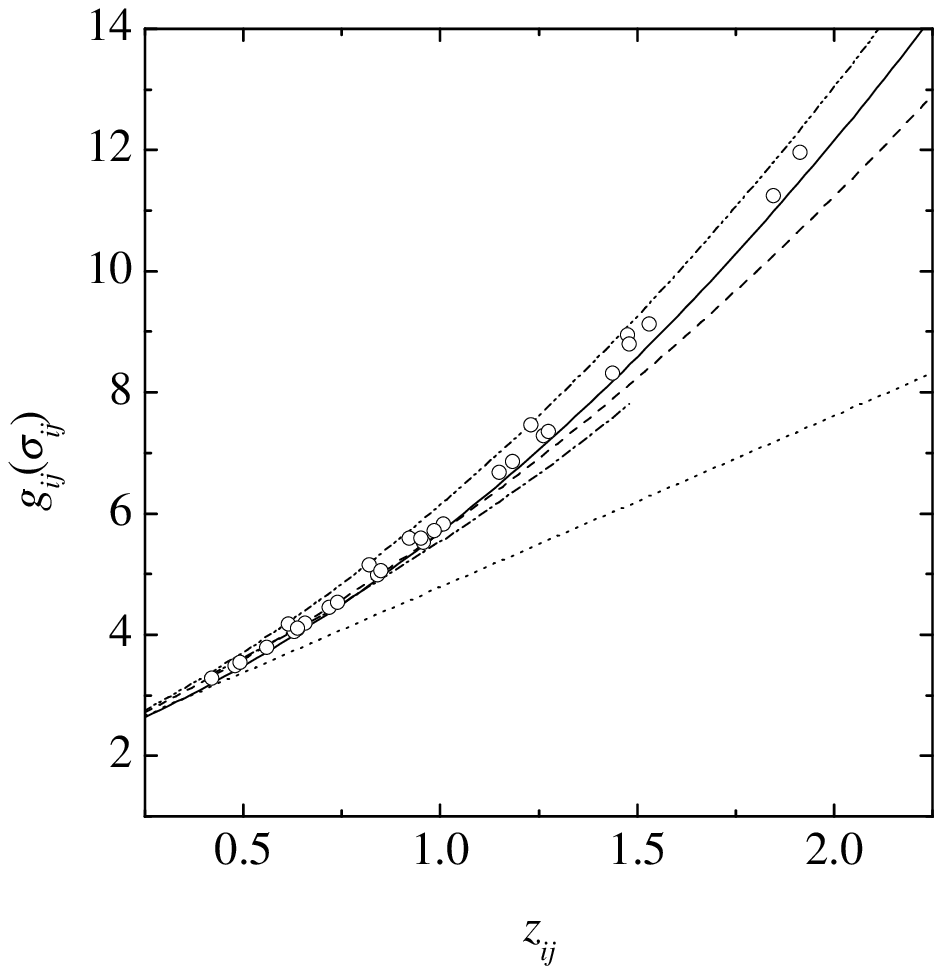}
\caption{{Plot of the contact value $g_{ij}(\sigma_{ij}^+)$ as a
function of the parameter
$z_{ij}=(\sigma_i\sigma_j)\langle\sigma^2\rangle/\langle\sigma^3\rangle$ for
ternary additive hard-sphere mixtures at a packing fraction $\eta=0.49$. The
circles are simulation data for the five cases A--E considered in this
paper. The lines are theoretical predictions: Eq.\ (\protect\ref{16}) (---),
BGHLL (-- -- --), SPT (-- $\cdot\cdot$ --), MS (-- $\cdot$ --), and PY
($\cdots$).}
\label{fig3}}
\end{figure}

Figures \ref{fig4}--\ref{fig8} show all the rdf $g_{ij}(r)$
($i,j=1,2,3$) as functions of the shifted distances $r-\sigma
_{ij}$ for the five different systems considered. Also included in
these figures are insets with an enlarged scale around
$g_{ij}(r)=1$ in which we have plotted $g_{ij}(r)$ versus the
actual rather than the shifted distance. It should be noted that,
{as said before,} the solution to the OZ equation with the MS
closure did not converge for cases C and E. 
This is a consequence of the fact that a term under the square root
in Eq.~(\ref{MS}) becomes negative at high densities; the authors of the MS
closure speculate that the lack of convergence may be a signal of phase
 transition \cite{MSconvergence}. The analysis of this 
conjecture is beyond the scope of this paper.
 The study of 
Figs.~\ref{fig4}--\ref{fig8} indicates the following. The RFA approach provides an
excellent overall agreement with the simulation results, which is
especially good in the region around contact. Something similar
occurs with the solution to the OZ equation with the MS closure,
except that this solution tends to underestimate the contact value
of $g_{23}$ and $g_{33}$. The PY closure clearly yields the
poorest results especially in the region around contact. All three
theoretical approaches lead to almost identical results beyond the
first minimum and exhibit a rich fine structure as was already
pointed out for another ternary system in Ref.\ \cite{BSL98}. The
fact that the simulation results also exhibit this structure is in
our view remarkable. It should be noted that there are slight
quantitative differences around the first minimum, which is more
pronounced in the theoretical solutions than in the simulation.
Except for cases C and E where the fine structure is rather
similar, in the other three cases the fine structure is case
dependent. As may be observed in Fig. \ref{fig2}, cases C and E
correspond to a situation where all partial packing fractions are
rather similar. Interestingly enough, when this happens, {\em
i.e.} no species is dominant with respect to volume occupation,
all the rdf $g_{ij}(r) $ relax to $1$ following an ordered
sequence of damped oscillations. Finally, it is also worth
mentioning that, for a given system, the form of the fine
structure of the $g_{ij}(r)$ is {\em almost\/ }the same for all
pairs. In fact, such  fine structure seems to evolve smoothly as
$\sigma _{ij}$ increases ($\sigma _{11}=1$, $\sigma _{12}=1.5$,
$\sigma _{22}=\sigma _{13}=2$, $\sigma _{23}=2.5$, $\sigma
_{33}=3$) as one can easily see by following the sequence top
pannel left , top pannel right, middle pannel right, middle pannel
left, bottom pannel right, bottom pannel left in Figs.\
\ref{fig4}, \ref{fig5}, and \ref{fig7}.
\begin{figure*}[h]
%[tbp]
\includegraphics[width=2 \columnwidth]
{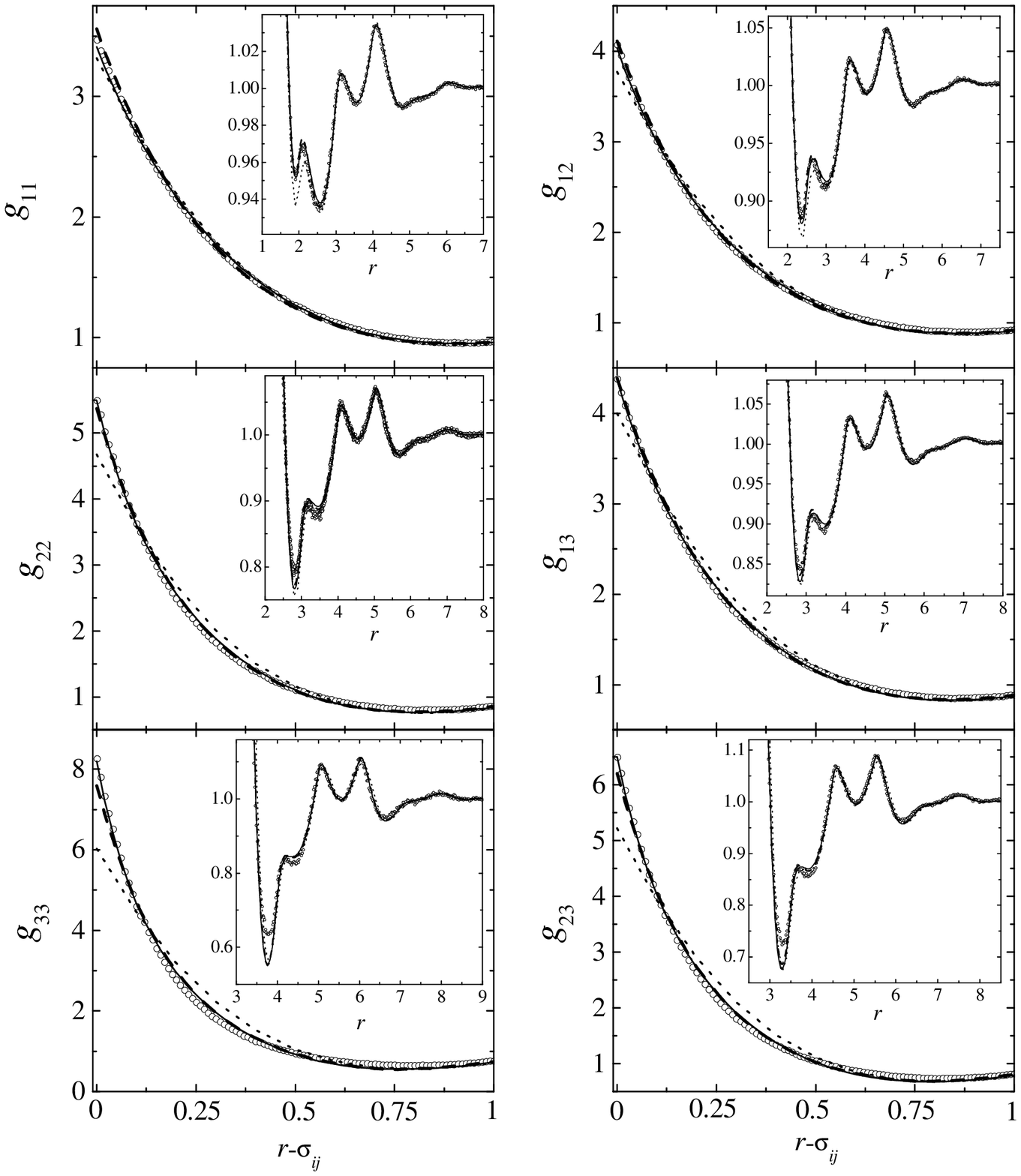}
\caption{{Radial distribution functions $g_{ij}(r)$ for a ternary mixture
with diamters $\sigma_1=1$, $\sigma_2=2$, $\sigma_3=3$ at a packing
fraction $\eta=0.49$ in case A ($x_1=0.7$, $x_2=0.2$, $x_3=0.1$). The
circles are simulation results, the solid lines are the RFA predictions, the
dotted lines are the PY predictions, and the dashed lines are the MS
predictions.} \label{fig4}}
\end{figure*}
\begin{figure*}[tbp]
\includegraphics[width=2 \columnwidth]
{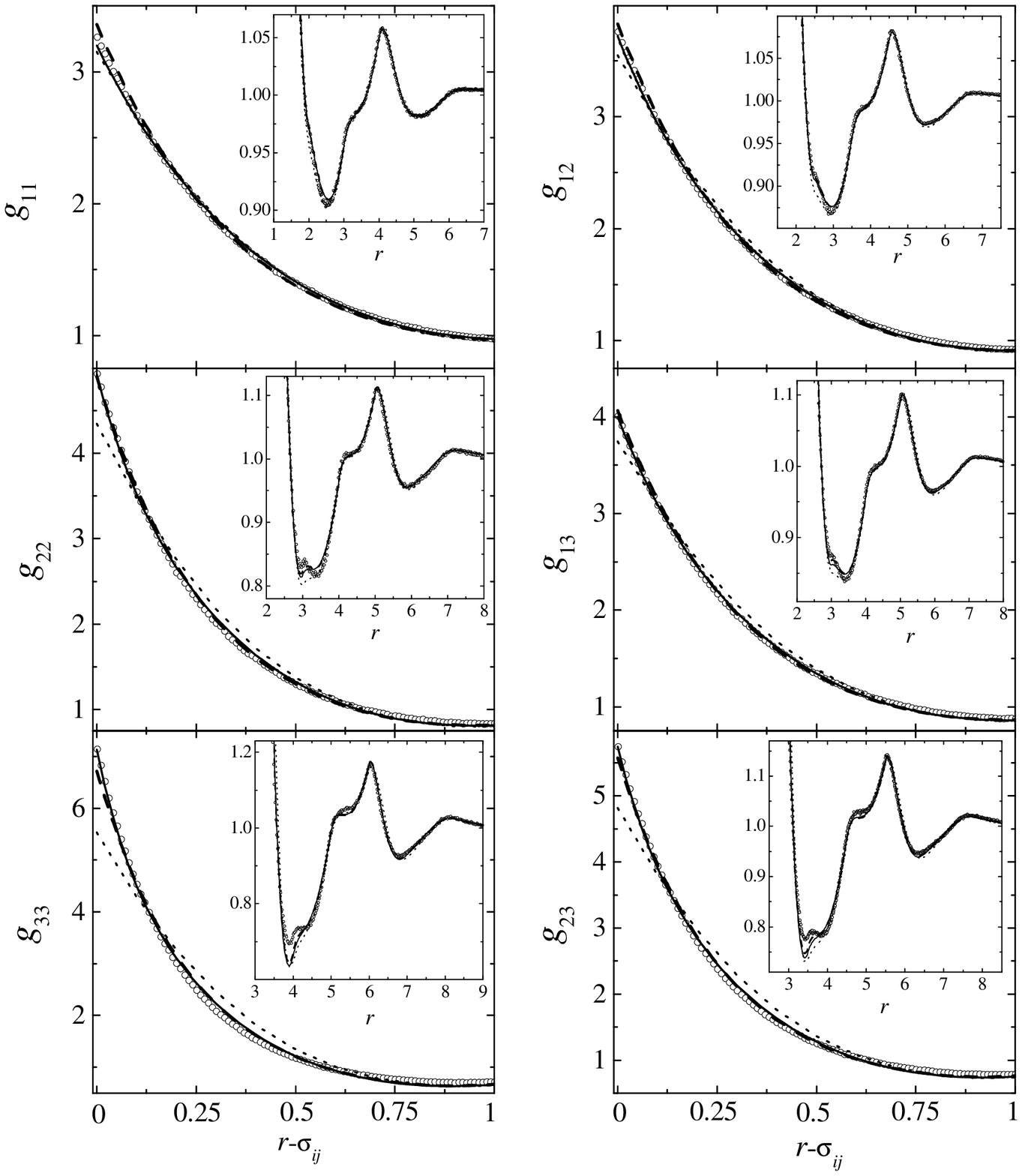}
\caption{{The same as in Fig.\ \protect\ref{fig4}, but in case B
($x_1=0.6$, $x_2=0.2$, $x_3=0.2$).} \label{fig5}}
\end{figure*}
\begin{figure*}[tbp]
\includegraphics[width=2 \columnwidth]
{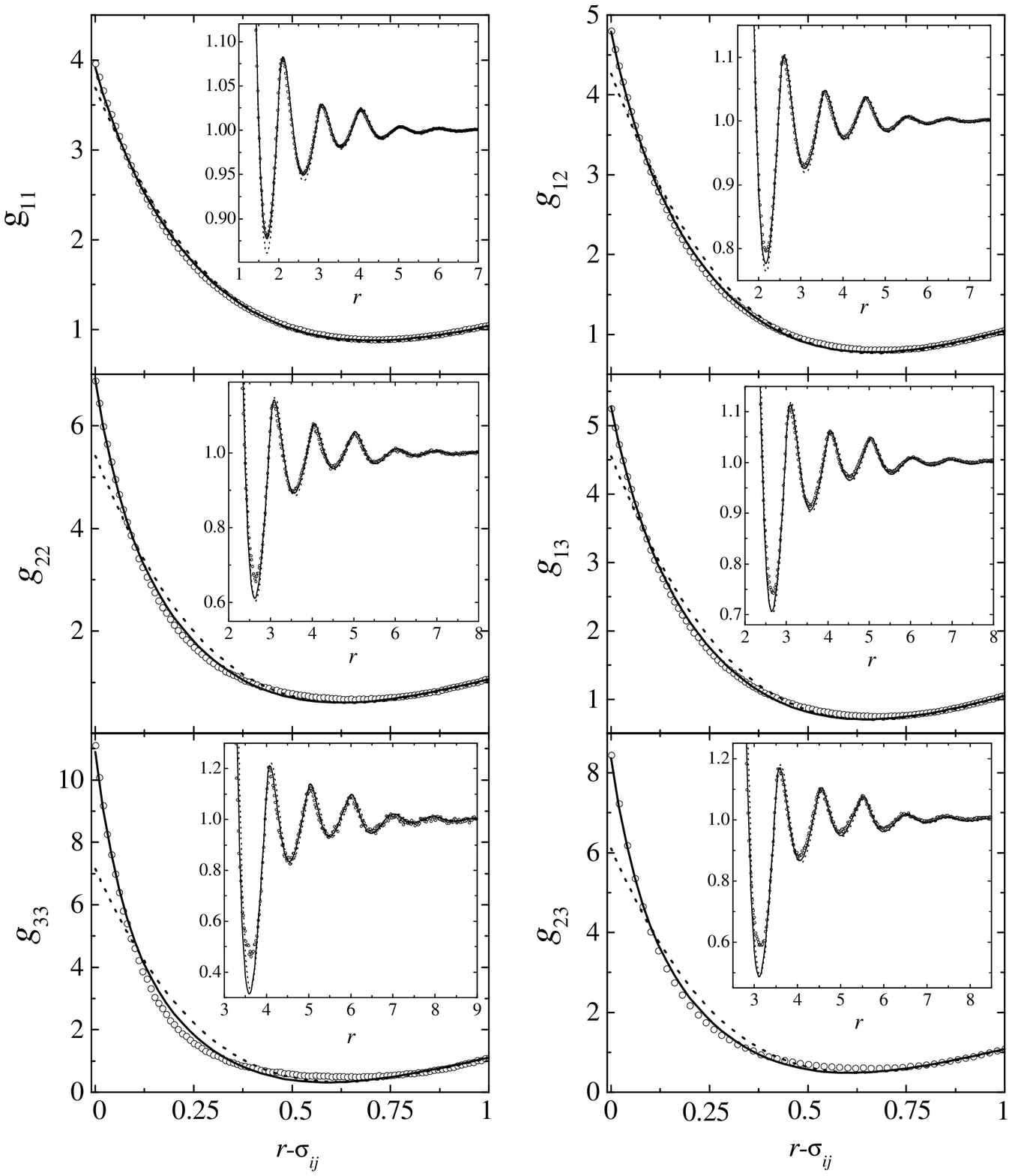}
\caption{{The same as in Fig.\ \protect\ref{fig4}, but in case C
($x_1=\frac{216}{251}$, $x_2=\frac{27}{251}$, $x_3=\frac{8}{251}$). Note
that the OZ equation with the MS closure fails to converge in this case.}
\label{fig6}}
\end{figure*}
\begin{figure*}[tbp]
\includegraphics[width=2 \columnwidth]
{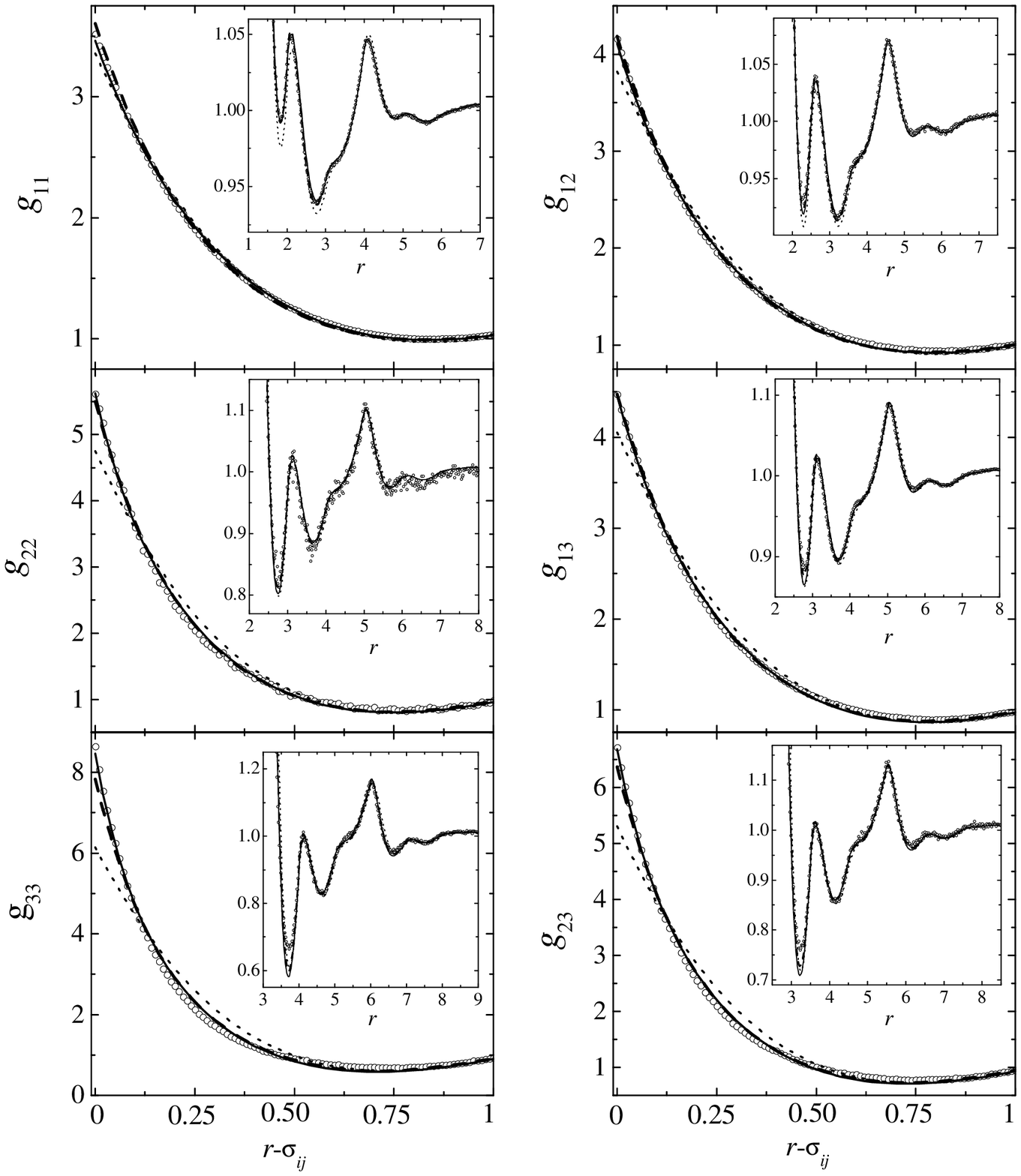}
\caption{{The same as in Fig.\ \protect\ref{fig4}, but in case D
($x_1=0.85$, $x_2=0.05$, $x_3=0.10$).} \label{fig7}}
\end{figure*}
\begin{figure*}[tbp]
\includegraphics[width=2 \columnwidth]
{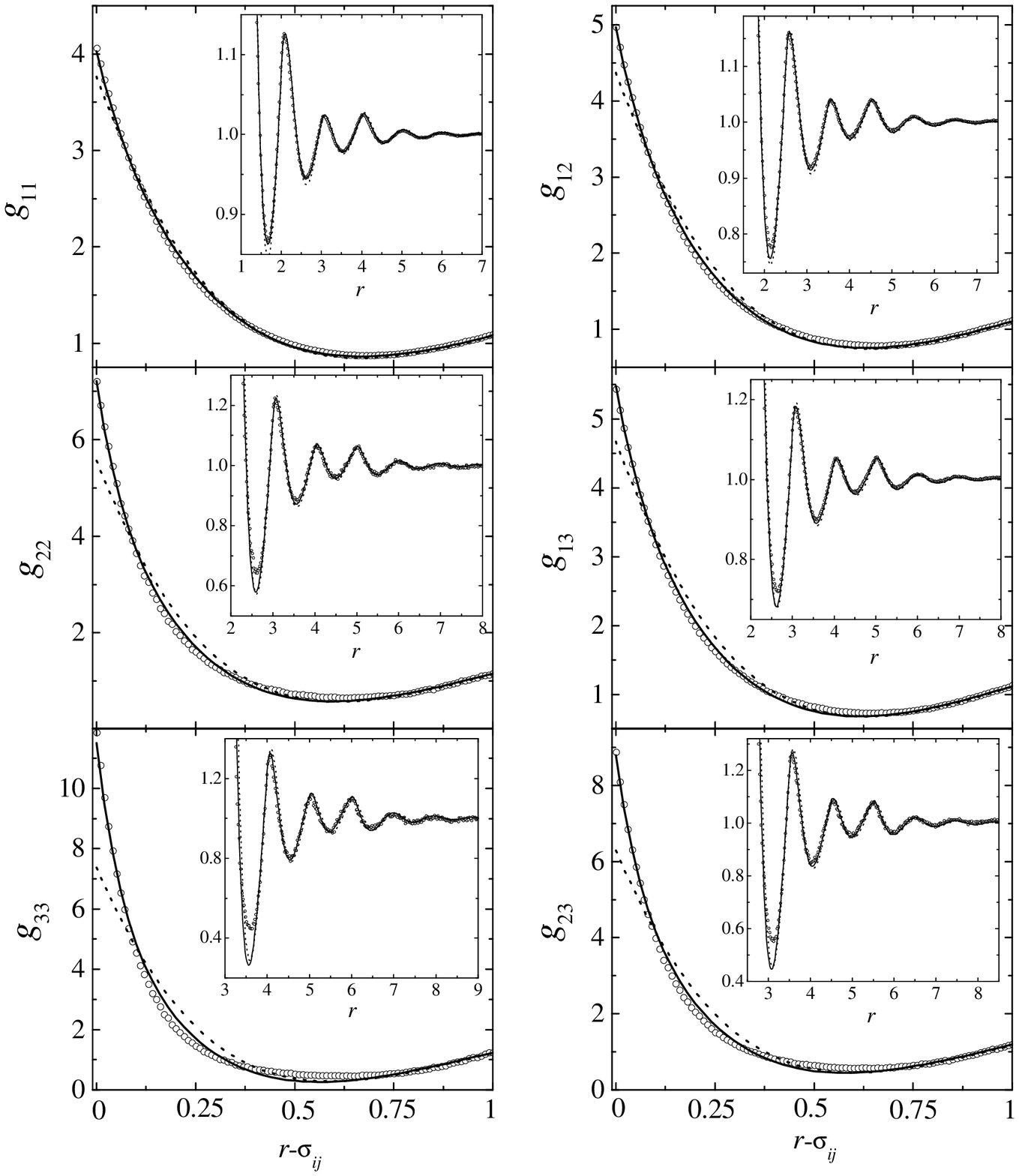}
\caption{{The same as in Fig.\ \protect\ref{fig4}, but in case E
($x_1=0.90$, $x_2=0.07$, $x_3=0.03$). Note
that the OZ equation with the MS closure fails to converge in this case.}
\label{fig8}}
\end{figure*}

\section{Concluding Remarks\label{sect4}}

The results of the previous section deserve some further comments.
As far as the simulations are concerned, we have here provided
further data on the thermodynamic and structural properties of
ternary additive hard-sphere fluid mixtures that extend and
complement those of Ref.\ \cite{bou2}. The quality and reliability
of these new data is reflected in their ability to capture the
rich fine structure that had been observed earlier in connection
with the RFA approach \cite{BSL98}. Both the RFA results and the
ones derived from the solution of the OZ equation with the MS
closure are in very good agreement with the simulation data, but
the latter give less accurate contact values,  the OZ equation
needs to be solved numerically, and  it presents convergence
problems when the partial packing fractions of all three species
have similar values. In any case these two theoretical approaches
do represent an improvement over the Percus--Yevick theory.
Finally, we have only carried out a preliminary qualitative
analysis of the rich fine structure that arose in the systems we
examined. By restricting ourselves to a fixed total packing
fraction and given diameter ratios, we attempted to investigate
the effect of partial volume occupation by each species on the
resulting structure. It thus appears interesting to assess the
effect of different total packing fractions and (or) size ratios.
We may address these and other related issues in multicomponent
systems in the future. \acknowledgments  The two first authors (A.
M. and A. M.) acknowledge support by the Center for Complex
Molecular Systems and Biomolecules (Project LN00A032 of the
Ministry of Education, Youth and Sports of the Czech Republic).
A.S. and S.B.Y. acknowledge partial support from the Ministerio de
Ciencia y Tecnolog\'{\i}a
 (Spain) through grant No.\ BFM2001-0718.
 We thank Prof. Lab\'{\i}k for helpful
discussions.

\end{document}